
%
%
%
\magnification=\magstep1
\overfullrule=0pt
\setbox0=\hbox{{\cal W}}

\def\w{{\cal W}}

\def\n{{\cal N}}

\def\ob{\bigl (}
\def\cb{\bigr )}
\def\lb{\lbrack}
\def\rb{\rbrack}
\def\de{\partial}

\def\q#1{\lb#1\rb}
\def\mn{\medskip\smallskip\noindent}
\def\sn{\smallskip\noindent}
\def\bn{\bigskip\noindent}

\font\extra=cmss10 scaled \magstep0 \font\extras=cmss10 scaled 750

\setbox1 = \hbox{{{\extra R}}}
\setbox2 = \hbox{{{\extra I}}}
\setbox3 = \hbox{{{\extra C}}}

\def\One{{{\extra 1}}\hskip-\wd3\hskip0.5 true pt{{\extra 1}}\hskip-\wd2
\hskip-0.5 true pt\hskip\wd3}
\def\id{\hbox{{\extra\One}}}
\setbox4=\hbox{{{\extra Z}}}
\setbox5=\hbox{{{\extras Z}}}
\setbox6=\hbox{{{\extras z}}}

\def\hwv{\mid \!h,w \rangle\, }
\def\ahwv{\langle h,w \! \mid\, }

\def\vac{\mid \!v \rangle\, }
\def\avac{\langle v \! \mid\, }
\def\onehalf{{\textstyle {1 \over 2}}}
\def\Commute{COMMUTE}
\def\commute{{\Commute} }
\def\bpz{1}
\def\felder{2}
\def\ginsparg{3}
\def\car{4}
\def\baf{5}
\def\bal{6}
\def\blg{7}
\def\nahm{8}
\def\nam{9}
\def\kau{10}
\def\blm{11}
\def\wirrep{12}
\def\rva{13}
\def\mfl{14}
\def\twistpap{15}
\def\supwir{16}
\def\supwirrep{17}
\def\zam{18}
\def\bouwknegt{19}
\def\gepnera{20}
\def\gepnerb{21}
\def\banks{22}
\def\pope{23}
\def\horst{24}
\font\HUGE=cmbx12 scaled \magstep4
\font\Huge=cmbx10 scaled \magstep4
\font\Large=cmr12 scaled \magstep3

\font\large=cmr17 scaled \magstep0
%
%
\nopagenumbers
\pageno = 0
\centerline{\HUGE Universit\"at Bonn}
\vskip 10pt
\centerline{\Huge Physikalisches Institut}
\vskip 2cm
\centerline{\Large A Note on the Algebraic Evaluation}
\vskip 6pt
\centerline{\Large \phantom{g} of Correlators in Local Chiral \phantom{g}}
\vskip 6pt
\centerline{\Large Conformal Field Theory}
\vskip 1cm
\centerline{\large A.\ Honecker}
\vskip 1.5cm
\centerline{\bf Abstract}
\vskip 15pt
\noindent
We comment on a program designed for the study of local chiral algebras
and their representations in 2D conformal field theory. Based on the
algebraic approach described by W.\ Nahm, this program efficiently
calculates arbitrary $n$-point functions of these algebras. The program
is designed such that calculations involving e.g.\ current algebras,
$\w$-algebras and $N$--Superconformal algebras can be performed.
As a non-trivial application we construct an extension of the Virasoro
algebra by two fields with spin four and six using the $N=1$--Super-Virasoro
algebra.
\vfill
\settabs \+&  \hskip 110mm & \phantom{XXXXXXXXXXX} & \cr
\+ & Post address:                       & BONN-HE-92-25   & \cr
\+ & Nu{\ss}allee 12                     & hep-th/9209029  & \cr
\+ & W-5300 Bonn 1                       & Bonn University & \cr
\+ & Germany                             & September 1992  & \cr
\+ & e-mail:                             & revised version & \cr
\+ & unp06b@ibm.rhrz.uni-bonn.de         & ISSN-0172-8733  & \cr
\eject
\pageno=1
\footline{\hss\tenrm\folio\hss}
%
\leftline{\bf 1.\ Introduction}
\mn
There are many different approaches to conformal field theory (CFT).
One can study the field algebra itself using e.g.\ the operator product
expansion (OPE) $\q{\bpz}$, one can examine the analyticity properties of the
correlation functions $\q{\felder}\q{\ginsparg}$, explore relations to
different fields in mathematics (e.g.\ modular invariance $\q{\car}$),
consider the conserved currents of Toda field theories
$\q{\baf}\q{\bal}\q{\blg}$ or one can --among others-- focus on the algebraic
aspects encoded in the mode algebra of the fields. In the latter approach
W.\ Nahm was able to derive a general \break
formula for the commutator of two quasiprimary local
chiral fields $\q{\nahm}$. Even more, he introduced a quasiprimary normal
ordering prescription $\n$ which in addition to the de\-riv\-a\-tive enables
one to construct all fields in the local chiral CFT from a few `simple' ones
(for more details see $\q{\nam}$). This approach lends itself for the
implementation in some computer algebra system. Very soon, it turned out
that one of the major computational tasks is the calculation of structure
constants, or more precisely the correlators involved $\q{\kau}\q{\blm}$.
\medskip
In order to make this more transparent let us give a very simple example.
Consider the `Virasoro algebra' which is defined in terms of generators
$L_n$ satisfying the following commutation relations:
$$\lb L_m,L_n\rb = (n-m)L_{m+n}+{c\over12}(n^3-n)\delta_{n+m,0}.
    \eqno({\rm 1.1})$$
$c$ is a central element in the algebra and in an irreducible
representation is equal to $C \id$ with $C$ a number. It is
convenient to identify the central element $c$ with the number $C$.
In physics, one very important representation of this algebra is the `vacuum
representation'. The vacuum representation is defined via the existence
of a cyclic vector $\vac$ that satisfies:
$$ L_n \vac = 0 \phantom{XXXXX} \forall n < 2. \eqno({\rm 1.2})$$
Correlators of the Virasoro algebra are now defined in terms of vectors
$L_{i_1} \dots L_{i_r} \vac$ and elements of the dual
space. The dual space can be characterized by a linear form $\avac$
dual to $\vac$:
$$\avac = (\vac)^{+} \eqno({\rm 1.3})$$
and an involution:
$$(L_n)^{+} = L_{-n}. \eqno({\rm 1.4})$$
A trivial consequence of (1.2) -- (1.4) is:
$$\avac L_n = 0 \phantom{XXXXX} \forall n > -2. \eqno({\rm 1.5})$$
Correlators of these states will now look as follows:
$$\avac L_{i_1} \dots L_{i_r} \vac = p(c) \eqno({\rm 1.6})$$
and will be evaluable as a (possibly trivial) polynomial in $c$.
\medskip
It is not difficult to implement the Virasoro algebra (1.1)
and its vacuum representation (1.2) -- (1.5) in REDUCE.
The following few lines will do the job:
\mn
{\tt \obeylines \obeyspaces
operator bra, ket, l;
noncom bra, ket, l; \mn
operator delta; \mn
\% the delta-operator \mn
for all m such that numberp m let delta(m) =
\hskip 40pt begin
\hskip 60pt if m=0 then return 1
\hskip 60pt else return 0
\hskip 40pt end; \mn
\% the Virasoro algebra itself \mn
for all m, n such that numberp m and numberp n and m<n let l(m)*l(n) =
\hskip 40pt l(n)*l(m)+(n-m)*l(n+m)+c/12*(n*n*n-n)*delta(n+m); \mn
\% the vacuum-representation of the Virasoro algebra \mn
for all n such that numberp n and n<2 let l(n)*ket(0,0) = 0;
for all n such that numberp n and n>-2 let bra(0,0)*l(n) = 0; \mn
let bra(0,0)*ket(0,0) = 1; \mn
}
\mn
In order to test the performance of this program we would like to calculate
to following correlator:
$$\avac L_{-N} \dots L_{-2} L_{-1} L_1 L_2 \dots L_N \vac
\phantom{XXXX} N \ge 1.  \eqno({\rm 1.7})$$
The appropriate REDUCE-input for such a correlator
reads -- specializing to $N=5$:
\mn
\centerline{\tt
bra(0,0)*l(-5)*l(-4)*l(-3)*l(-2)*l(-1)*l(1)*l(2)*l(3)*l(4)*l(5)*ket(0,0);
}
\mn
We have evaluated the first few correlators of this form using REDUCE 3.3
on a IBM 3084 mainframe. The CPU-time needed (in seconds) is listed in the
first column of the following table:
\mn
\centerline{
\vrule \hskip 1pt
\vbox{ \offinterlineskip
\def\tablespace{ height2pt&\omit&&\omit&&\omit&&\omit&&\omit&\cr }
\def\tablerule{ \tablespace
                \noalign{\hrule}
                \tablespace        }
\hrule
\halign{&\vrule#&
  \strut\hskip 4pt\hfil#\hfil\hskip 4pt\cr
\tablespace
& $N$ && REDUCE 3.3 && \commute  && REDUCE 3.4   && \commute    &\cr\tablespace
& \omit && IBM 3084 && IBM XT286 && DEC 5000/120 && DEC 5000/120&\cr\tablespace
\noalign{\hrule}
\tablerule
& $1$ && $0.43$     &&  $< 1$    && $0.1$        && $0.0$       &\cr\tablespace
& $2$ && $0.77$     &&  $< 1$    && $0.1$        && $0.0$       &\cr\tablespace
& $3$ && $1.34$     &&  $< 1$    && $1.0$        && $0.0$       &\cr\tablespace
& $4$ && $9.34$     &&   $1$     && $8.4$        && $0.0$       &\cr\tablespace
& $5$ && $364.42$   &&   $16$    && $185.2$      && $0.7$       &\cr\tablespace
& $6$ && \omit      &&  $448$    && $14000.9$    && $21.8$      &\cr\tablespace
& $7$ && \omit      &&  $18114$  && \omit        && $895.3$     &\cr\tablespace
& $8$ && \omit      &&  \omit    && \omit        && $50735.4$   &\cr\tablespace
}
\hrule}\hskip 1pt \vrule
}
\mn
%
Missing entries in the table indicate that completion of the program could
not sensibly achieved with the existing resources, e.g.\ a CPU-time limit
of 30min on the mainframe.
\sn
Obviously, performance of the above program is poor and calculation can be
much more efficiently performed on a machine as small as a IBM XT286 running
at 6MHz CPU clock. We will refer to this program written by the author as
`\Commute'. A better comparison of both programs can be derived from
CPU time needed by both programs on a DEC workstation
\footnote{${}^{1})$}{
A SUN 4/75 sparc station at 40MHz is about 15\% faster.
}.
Note that we have chosen the correlator (1.7) because it is worst case
for \commute (see also chapter 3).
\mn
This letter is to a great deal dedicated to the algorithmic
considerations implemented in \Commute. Since they have been developed
while performing calculations the next section intends to give a motivation
for some of the implementation's pecularities through a short historic review.
We will further outline the capabilities
of \commute and mention some possible applications.
Finally, as a non-trivial application of the program we will explicitly
show that the structure constants in the bosonic sector of the $N=1$--Super
Virasoro algebra coincide with those of $\w(2,4,6)$.
\medskip
{}From now on we will use notations and formulae introduced
in chapter 2 of $\q{\blm}$ as well as chapter 3 of $\q{\wirrep}$
and assume that the reader is familiar with them.
\bn
\leftline{\bf 2.\ History of the program}
\mn
In $\q{\blm}$ a PASCAL program was used for the evaluation of the most
complicated structure constants. First calculations for representations of
$\w$-algebras $\q{\rva}\q{\mfl}$ had also reached the limits of
REDUCE. The analogue of the REDUCE program presented in chapter 1
implemented in MATHEMATICA was even slower.
Thus it was clear that for explicit calculations with $\w$-algebras a much
more efficient program was needed.
Before going on with representation theory of $\w$-algebras $\q{\wirrep}$
we therefore decided to develop a special
purpose program in C. This had the disadvantage that even basic math routines
for large integers had to be written again but optimal control on rule
application was guaranteed.
\sn
The first version of \commute was designed only to implement evaluation of
correlators as presented in chapter 1. Soon, it turned out that expansion
of normal ordered products produced so many summands that even their
evaluated form could not be read any more by the computer algebra systems
we were using. This lead to the implementation of mode's and expansion
of naive normal ordered products in \Commute. Not much later also expansion
of quasiprimary normal ordered products into naive normal ordered products
blew up expressions so much that it had to be implemented in \Commute.
As only for $\n(X,\de^n L)$ a simple formula exists $\q{\wirrep}$ it was
necessary to tell \commute how to evaluate the other $\n$'s. This was done
introducing a file containing definitions. Now,
only a few minor changes were necessary to also make calculations with
twisted representations of
bosonic $\w$-algebras (where some of the bosonic fields have half-integral
modes $\q{\twistpap}$) possible.
\sn
Next, we turned to $N=1$--${\cal SW}$-algebras $\q{\supwir}$. Now the
$N=1$--Super-Virasoro algebra had to be implemented in \commute. Only a few
minor improvements were necessary until representations of higher spin
${\cal SW}$-algebras in the Ramond-sector needed special attention
$\q{\supwirrep}$. At this stage \commute was already
much more powerful than originally intended
and performed most of the tasks earlier performed in REDUCE or MATHEMATICA.
This motivated us to completely make it a stand-alone system. To this end
the user had to be enabled to introduce new fields and their commutators.
Therefore Nahm's universal polynomials had to be implemented at last and it
seemed prudent to implement replacement rules for structure constants and
insertion of the central charge $c$.
\medskip
It also turned out while developping large programs in REDUCE that its
tendency to accept almost any input and almost never to produce any
error messages made debugging of these programs difficult.
Even more, the error condition is almost impossible to locate due to
a missing trace-option. Therefore we implemented error checking in
\commute from the beginning. Even in an error condition \commute continues
and tries to make sense of what follows. Additionally, \commute also
supports a debugging mode in which it displays all steps it performs.
This enables the user to control proper performance of \commute and
faciliates debugging.
\medskip
This historical review should motivate some of the pecularities of \Commute.
Though \commute can do much on its own,
the use of some computer algebra system is recommended e.g.\ for factorization
of polynomials and evaluation of the explicit form of complicated $\n$opp's.
\bn
\leftline{\bf 3.\ Algorithms used}
\mn
First, we would like to make some comments that should be generally
valid for computer algebra systems when applied to noncommutative objects.
The first observation is that commutative and noncommutative objects should
be stored in different ways because their treatment is completely
different. We have chosen to store correlators as a vector of structures
describing the modes of fields. This faciliates identification of
locations where commutation should take place.
Storing the polynomial in front
of it is conceivable in many different ways. For our purposes it seemed
sensible to store it as
an array with integer coefficients of monoms in $h$ and ${c \over 12}$ with
a common denominator and a common monom in structure constants.
\mn
It is obvious that rules which define vanishing terms simplify expressions
and should be applied as often as possible. Therefore, only one
commutator resp.\ one field inserted in a step.
Afterwards, those correlators are eliminated that vanish because the mode
of the field annihilates the highest weight. Term
collection also simplifies expressions but for general correlators is very
time consuming
and therefore seldom performed. Still, summing up polynomials is a comparably
simple task and performed as often as possible.
It should be possible to implement some heuristic method to keep expressions
small in most computer algebra systems. The author does in fact not understand
why (as far as he knows) no one of the computer algebra systems mentioned above
tries to collect terms
as soon as memory becomes rare. Such a simple improvement would certainly
make tasks manageable that are currently impossible.
\medskip
Let us now focus on algorithms that are more specific to conformal field
theory. In our notation for correlators all vectors are $L_0$-eigenvectors:
$$L_0 \left( \phi_{1,i_1} \dots \phi_{r,i_r} \hwv \right)
  = (h+i_1+ \dots i_r) \phi_{1,i_1} \dots \phi_{r,i_r} \hwv.
\eqno({\rm 3.1})$$
Thus it is very simple to eliminate any $L_0$ and \commute performs this at
the same time when checking for vanishing correlators.
\sn
If a product of two same modes of the same fermion is encountered (this does
generally not vanish) it can be replaced by one half of the anticommutator.
This also is a considerable simplification and therefore performed next.
\sn
Even if we really have to evaluate a correlator there is a strategy that
performs much better than the REDUCE-program described in chapter 1. First
note that if you can evaluate $\lb L_{-r}, L_r \rb$, it is almost as simple
to evaluate $\lb L_{-r}, L_r^n \rb$. \commute indeed does this and therefore
a correlator of the form (1.7) really is the worst case for \commute -- e.g.\
computation of $\ahwv L_{-2}^{10} L_2^{10} \hwv$ is performed instantly.
\sn
As we would like to perform no term collection it seems prudent to finish
any summand currently being processed before proceeding with the next one.
In most cases only a few steps are necessary to produce a scalar expression.
In the meanwhile a few additional summands are produced inserting commutators.
These are processed next. For a few exceptional cases, this strategy may be
far from optimal as far as CPU-time is concerned. Definitely, it will require
a minimum of memory. Within this strategy it is obviously most efficient
to produce as few additional summands as possible before arriving at a
scalar expression. This is ensured by searching the place most close
to either $\ahwv$ or $\hwv$ for possible insertion of a commutator
and performing it there instead of doing so e.g.\ at the leftmost possible
position.
\medskip
The remaining features of \commute are implemented rather heuristically.
Expansion of modes of fields is performed before any commutation is performed
with the restriction that only one naive normal ordered product is expanded
at a time because this blows expressions up. Externally defined commutators
are inserted if no one of the internally defined ones applies any more.
Replacement rules on structure constants are applied at the same time
when term collection takes place. Other features of \commute like replacement
of quadratic fields in the Ramond-sector are also performed with low
priority.
\bn
\leftline{\bf 4.\ Availability, use and applications}
\mn
\commute is publically available together with a detailed technical
description which we omit here. It may be redistributed and changed
freely if the conditions supplied with the program are respected.
\mn
\commute is written entirely in ANSI-C. Therefore it should run on almost every
PC or workstation. Before compilation, compatibilty to either REDUCE- or
MATHEMATICA-notation can be chosen. Additionally, some limits (especially
numerical range) may be specified before compilation so that
many different versions of
\commute are conceivable. Therefore we have decided to distribute appropriate
versions on request via e-mail to the author rather than
to put the program on an internet-server.
\mn
Anybody interested in \commute should send an e-mail to
{\tt unp06b@ibm.rhrz.uni-bonn.de} specifying the computer being used
and the version of \commute needed. You shoud also specify whether you would
like a compiled version and/or the sources. The desired medium
(ftp / disc / tape) should also be specified.
\medskip
\commute is invoked from the command line and replaces any correlator
in an input file by its value producing an output file. This procedure
makes it possible to call \commute from most high level languages.
\medskip
Some example files for what \commute can do without any other computer
algebra system (except for factorization of polynomials) are included
with it. They cover e.g.:
\item{$\bullet$} The associativity check for four point correlators
                 of the algebra $\w(2, {5 \over 2})$ reproducing the
                 result of R.\ Varnhagen that only finitely many HWRs
                 are permitted $\q{\rva}$.
\item{$\bullet$} Evaluation of correlators involving the simplest null
                 fields in the
                 Super-Virasoro algebra at $c={7 \over 10}$ yielding
                 the permitted $h$-values of the minimal series in the
                 Neveu-Schwarz- and Ramond-sector. Special attention is
                 paid to the quadratic fermionic normal ordered product
                 in the Ramond-sector
                 where the projection on the quasiprimary part according
                 to W.\ Nahm $\q{\nahm}$ is impossible due to the non-local
                 effect of the boundary conditions.
\item{$\bullet$} Definition of Zamolodchikov's $\w(2,3)$ $\q{\zam}$. This
                 definition needs only very few lines, but enables one to
                 calculate e.g.\ the contravariant form on the Verma module
                 explicitly, thus demonstrating the efficieny of \Commute.
\item{$\bullet$} Definitions for the $N=2$--Super-Virasoro-Algebra.
\bn
\leftline{\bf 5.\ An application:
The bosonic sector of the $\bf N=1$--Super-Virasoro algebra}
\mn
P.\ Bouwknegt $\q{\bouwknegt}$ has argued some time ago that the bosonic sector
of the $N=1$--Super-Virasoro algebra should be an extension of the
Virasoro algebra with one field of spin 4 and one field of spin 6 --
called a $\w(2,4,6)$.
We shall now verify this and calculate the structure constants.
\mn
The $N=1$--Super-Virasoro algebra is given by the following commutation
relations:
$$ \eqalign{ \lb L_m,L_n\rb &= (n-m)L_{m+n}+{c\over12}(n^3-n)\delta_{n+m,0} \cr
             \lb L_m,G_n\rb &= (n-\onehalf m)G_{m+n} \cr
             {\lb G_m,G_n\rb}_{+} &= 2 L_{m+n}
              + {c\over3}(m^2-{\textstyle {1\over4}})\delta_{m+n,0}.
            } \eqno( {\rm 5.1}) $$
This algebra is already implemented in \Commute.
\sn
We are now interested in bosonic normal ordered products built up from
the field $G$ only. For dimension up to 6 there are only two of them:
$$\eqalign{\n(G, \de G) =& N(G,\de G)
                          - {1\over2} \de N(G,G)
                          + {1\over5} \de^2 L \cr
           \n(G, \de^3 G) =& N(G,\de^3 G)
                            - {3\over2} \de N(G, \de^2 G)
                            + {2\over3} \de^2 N(G, \de G)
                            - {1\over12} \de^3 N(G,G)
                            + {1 \over84} \de^4 L \cr
} \eqno( {\rm 5.2}) $$
where we have already expressed the quasiprimary normal ordered product `$\n$'
in terms of naive normal ordered products `$N$'.
Note that the normal ordered products (5.2) are not implemented in \commute
and therefore have to be defined in a definition file.
\sn
In order to construct primary
fields of dimension 4 resp.\ 6 one now takes arbitrary linear combinations
of all quasiprimary fields with this dimension and adjusts the parameters
such that structure constants become zero.
\sn
For a primary spin four field $U$ the most general ansatz is:
$$ U := \alpha \n(L,L) + \beta \n(G, \de G) . \eqno( {\rm 5.3} )$$
The free parameters $\alpha$ and $\beta$ have to be adjusted such that
$C_{U L}^L = 0$ holds. This yields the equation:
$$0 = C_{U L}^L = \alpha C_{\n(L,L) L}^L + \beta C_{\n(G, \de G) L}^L
                 = \alpha {5 c + 22 \over 5} + \beta {17 \over 5} .
\eqno( {\rm 5.4})$$
The structure constants $C_{X Y}^{Z}$ are defined as a solution of
the system of equations:
$$C_{X Y}^{W} d_{W Z} = C_{X Y Z} . \eqno({\rm 5.5})$$
with
$$\eqalign{
d_{X Y} &= \avac X_{-d(X)} Y_{d(Y)} \vac \cr
C_{X Y Z} &= \avac Z_{-d(Z)} X_{d(Z) - d(Y)} Y_{d(Y)} \vac . \cr
} \eqno({\rm 5.6})$$
The evaluation of the correlation functions in (5.6) is performed
using \commute while the equations (5.5) are afterwards easily solved using
e.g.\ REDUCE.
\sn
In (5.4) the result of the corresponding calculations has already been
inserted.
\sn
The general solution of (5.4) is $\alpha=\gamma {17 \over 5}$ and
$\beta = - \gamma {5 c + 22 \over 5}$ with a free parameter $\gamma$
which we can use for normalization. We choose $\gamma$ such that
$$\eqalign{
{c \over 4} =& \avac U_{-4} U_4 \vac \cr
            =& \alpha^2 \avac \n(L,L)_{-4} \n(L,L)_4 \vac \cr
             & + 2 \alpha \beta \avac \n(G,\de G)_{-4} \n(L,L)_4 \vac \cr
             & + \beta^2 \avac \n(G,\de G)_{-4} \n(G, \de G)_4 \vac . \cr
} \eqno( {\rm 5.7})$$
If we now insert the two-point-functions that we have already evaluated
using \commute we arrive at:
$$\gamma^2 = {75 \over 2 (10 c - 7 ) (5 c + 22 ) (4 c + 21)}.
\eqno( {\rm 5.8})$$
The most general ansatz for a primary field $V$ with dimension six is:
$$V := \kappa \n(G \de^3 G) + \lambda \n(\n(G, \de G), L)
                            + \mu \n(\n(L,L),L) + \nu \n(L, \de^2 L).
\eqno( {\rm 5.9})$$
Again one calculates structure constants using \Commute.
The first observation is that $C_{V L}^{\n(\n(G, \de G), \de L)} = 0$
holds independently of the free parameters $\kappa$, $\lambda$, $\mu$ and
$\nu$. Now there remain three non-trivial equations:
$$C_{V L}^L = C_{V L}^{\n(L,L)} = C_{V L}^{\n(G, \de G)} = 0 .
\eqno( {\rm 5.10})$$
Similar arguments as for the spin four field lead to the general solution:
$$\eqalign{ \mu &= 10 (218 c - 293) \rho \cr
            \nu &= - 3 (11 c - 86) (c + 24) \rho \cr
            \lambda &= -{5 \over 15} \ob {15 c + 164 \over 10} \mu
                                        + {16 \over 3} \nu \cb \cr
            \kappa &= - {c + 24 \over 26} \lambda \cr
} \eqno( {\rm 5.11})$$
with a free parameter $\rho$. Imposing the nomalization condition
$$ {c \over 6} = \avac V_{-6} V_6 \vac, \eqno( {\rm 5.12})$$
fixes $\rho^2$ to the following value:
$$\rho^2 = {1 \over 25 (14 c + 11) (10 c - 7) (7 c + 68) (4 c + 21)
                       ( 2 c - 1) (c + 24) ( c+ 11)} . \eqno( {\rm 5.13})$$
So far, no time intensive calculations need to be done. The next step
is to replace the field $\n(G, \de G)$ by the field $U$ and the field
$\n(G, \de^3 G)$ with $V$ and calculate again the $d$-matrix and structure
constants.
Especially the evaluation of the structure constants involving fields of
dimension up to seven took about a day's CPU-time on a SUN-SPARC
at the Max-Planck-Institut f\"ur Mathematik in Bonn-Beuel.
The matrix-inversion is easily performed using REDUCE.
\sn
The result shows that $U$ and $V$ are not only indeed primary but orthogonal
to all normal ordered products not involving $G$. This means that in
the bosonic sector of the Super-Virasoro algebra the fields $U$ and $V$ are
simple (i.e.\ non-composite). The non-trivial structure constants connecting
three simple fields are evaluated as:
$$\eqalign{
\ob C_{U U}^U \cb^2 &= { 54 (10 c^2 + 47 c - 82)^2 \over
                                  (10 c - 7) (5 c + 22) (4 c + 21)} \cr
\ob C_{U U}^V \cb^2 &= { 144 (14 c + 11) (5 c + 22)^2 (c+11) (c - 1)^2
                         \over
                         (10 c - 7) (7 c + 68) (4 c + 21) (2 c - 1) (c + 24)
                         } \cr
\ob C_{V V}^U \cb^2 &= { 50 (7 c + 68)^2 (3 c +20)^2 (2 c - 1)^2
                         \over
                         3 (10 c - 7) (5 c + 22) (4 c + 21) (c + 24)^2
                         } \cr
\ob C_{V V}^V \cb^2 &= { 400 (616 c^4 + 19106 c^3 + 183931 c^2 + 574876 c
                                      + 724096)^2 (c-1)^2
                         \over
                         9 (14 c +11) (10 c - 7) (7 c +68) (4 c + 21) (2 c - 1)
                           (c + 24)^3 (c + 11)
                         } . \cr
} \eqno( {\rm 5.14})$$
We have given only expressions for the squares in order to make it possible to
insert the normalization constants $\gamma^2$ and $\rho^2$. One obtains two
additional relations which fix signs among the coupling constants
$$\eqalign{
C_{U U}^U C_{V V}^U &= {30 (10 c^2 + 47 c -82) (7 c + 68) (3 c + 20) (2c-1)
                 \over (10c-7) (5c+22) (4c+21) (c+24) } \cr
C_{U U}^V C_{V V}^V &= {80 (616 c^4 + 19106 c^3 + 18391c^2 + 574876 c
                           + 724096) (5 c + 22) (c-1)^2
                 \over (10c-7) (7 c + 68) (4c+21) (2 c -1) (c+24)^2 } \cr
} \eqno( {\rm 5.15})$$
so that we are left with two free signs which reflects the freedom of choice
of signs for $\rho$ and $\gamma$.
\sn
Equalities like $C_{X V}^U = {2 \over 3} C_{X U}^V$ enable one to calculate
all remaining structure constants connecting three simple fields. If one now
calculates the dimension of the space of bosonic fields in the
$N=1$--Super-Virasoro algebra and for a $\w(2,4,6)$-algebra one observes
equality for integer dimension up to $9$. At dimension 10 there is
one field less in the bosonic projection of the $N=1$--Super-Virasoro algebra
compared to a general $\w(2,4,6)$-algebra. This shows that the fields
$L$, $U$ and $V$ generate a closed $\w(2,4,6)$ subalgebra in the bosonic
sector of the Super-Virasoro algebra. Since the $N=1$--Super-Virasoro algebra
satisfies Jacobi identities, so does this algebra, too.
\sn
There have been earlier constructions of $\w(2,4,6)$-algebras starting
with simple fields of dimension two, four and six. Imposing some Jacobi
identities H.G.\ Kausch and G.M.T.\ Watts have found two general solutions
$\q{\kau}$, but have not been able to check all relevant Jacobi
identities.
The structure constants (5.14), (5.15)  of the algebra $\w(2,4,6)$ we have
constructed here coincide with those of one of the two general solutions.
\bn
The cases where the normalization constants as well as
the structure constants become singular deserve special attention.
For $c \in \{ -{14 \over 11}, -{7 \over 68}, {1 \over 2}, -24, -11 \}$
the field $V$ is a null field before normalization and the procedure here
only leads to a $\w(2,4)$-algebra. For $c \in \{ {7 \over 10}, -{21 \over 4}
\}$ also the field $U$ is a null field before normalization and the bosonic
sector of the Super-Virasoro algebra coincides with the Virasoro algebra.
It might seem that for $c=-{22 \over 5}$ we obtain a $\w(2,6)$ which is not
the case.
Here, singularities in the structure constants $C_{V V}^{X}$ force us to also
normalize $V$ to zero.
This shows that there is no consistent
$\w$-algebra in the bosonic sector at $c=-{22 \over 5}$ and one is left
with the Virasoro algebra again.
\bn
It has been observed in $\q{\wirrep}$ that the representation theory of
$\w(2,4)$ and the $N=1$--Virasoro algebra at $c=-11$ is much the same. This
observation obviously should generalize to the bosonic sector of the
$N=1$--Virasoro algebra at arbitrary central charge.
\bn
\leftline{\bf 6.\ Conclusions}
\mn
We have described a program for the evaluation of correlators in local chiral
conformal field theory. It is fast, easy to use and pays
attention to wrong input. This program has already been used for representation
theory of $\w(2, \delta)$-algebras $\q{\wirrep}$, the construction of
$N=1$--${\cal SW} ({3 \over 2},d)$- and
${\cal SW} ({3 \over 2}, d_1, d_2)$-algebras $\q{\supwir}$ and finally for
the representation theory of ${\cal SW} ({3 \over 2},d)$ $\q{\supwirrep}$.
Now it has been extended in order to make calculations involving even more
general local chiral algebras possible. This includes $N$-extended
superconformal
algebras as well as current algebras. The next application will be
$N=2$--Super-$\w$-algebras because of their important applications in
String theory $\q{\gepnera}\q{\gepnerb}\q{\banks}$.
\sn
We have demonstrated the power of the program by constructing a $\w(2,4,6)$
starting from the $N=1$--Super-Virasoro algebra. Recently, H.\ Lu and
C.N.\ Pope have observed that any free field construction may start with
specific $\w$-algebras as input $\q{\pope}$.
\commute would be especially well suited to implement this for more algebras.
\bn
\bn
It is a pleasure to thank everybody working in the group of W.\ Nahm at the
`Physikalisches Institut' in Bonn for numerous valuable discussions. I am
especially grateful to W.\ Eholzer and R.\ H\"ubel because the original
concept of \commute was developed together with them. Furthermore, I am
indebted to the `Max-Planck-Institut f\"ur Mathematik' in Bonn-Beuel because
without their computers complicated applications for \commute
(like the one in this letter) would not have been possible.
\bn
\bn
{\bf Note added:} I would like to thank H.G.\ Kausch for pointing out to me
that he has also performed an explicit contruction of $\w(2,4,6)$ in terms
of the $N=1$--Super Virasoro algebra using a slightly different method
$\q{\horst}$.
\vskip 2.5cm
\leftline{\bf References}
\mn
\settabs\+&\phantom{---------}&\phantom
{------------------------------------------------------------------------------}
& \cr
\+ &$\q{\bpz}$ & A.A.\ Belavin, A.M.\ Polyakov, A.B.\ Zamolodchikov & \cr
\+ &           & {\it Infinite Conformal Symmetry in Two-Dimensional
                 Quantum Field Theory}  & \cr
\+ &           & Nucl.\ Phys.\ {\bf B241} (1984) p.\ 333  & \cr
\+ & $\q{\felder}$
               & G.\ Felder, J.\ Fr\"ohlich, G.\ Keller,
                 {\it Braid Matrices and Structure Constants for} & \cr
\+ &           & {\it Minimal Conformal Models}, Commun.\ Math.\ Phys.\ 124
                 (1989) p.\ 647 & \cr
\+ & $\q{\ginsparg}$
               & P.\ Ginsparg, {\it Applied Conformal Field Theory} & \cr
\+ &           & Conference Les Houches
                 `Fields, Strings and Critical Phenomena' (1990) & \cr
\+ &$\q{\car}$ & J.L.\ Cardy, {\it Operator Content of Two-Dimensional
                 Conformally Invariant Theories} & \cr
\+ &           & Nucl.\ Phys.\ {\bf B270} (1986) p.\ 186 & \cr
\+ &$\q{\baf}$ & J.\ Balog, L.\ Feh\'er, L.\ O'Raifeartaigh, P.\ Forg\'acs,
                 A.\ Wipf & \cr
\+ &           & {\it Toda Theory and $\w$-Algebra from a Gauged WZNW Point of
                 View} & \cr
\+ &           & Ann.\ Phys.\ 203 (1990) p.\ 76 & \cr
\+ &$\q{\bal}$ & J.\ Balog, L.\ Feh\'er, P.\ Forg\'acs, L.\ O'Raifeartaigh,
                 A.\ Wipf & \cr
\+ &           & {\it Kac-Moody Realization of $\w$-Algebras},
                 Phys.\ Lett.\ {\bf B244} (1990) p.\ 435 & \cr
\+ &$\q{\blg}$ & A.\ Bilal, J.L.\ Gervais & \cr
\+ &           & {\it Systematic Construction of Conformal Theories with
                 Higher-Spin Virasoro} & \cr
\+ &           & {\it Symmetries}, Nucl.\ Phys.\ {\bf B318} (1989) p.\ 579 &\cr
\+ &$\q{\nahm}$
               & W.\ Nahm, {\it Chiral Algebras of Two-Dimensional Chiral
                 Field Theories and Their} & \cr
\+ &           & {\it Normal Ordered Products}, Proceedings Trieste
                 Conference on & \cr
\+ &           & Recent Developments in Conformal Field Theories, ICTP,
                 Trieste (1989) p.\ 81 & \cr
\+ &$\q{\nam}$ & W.\ Nahm, {\it Conformal Quantum Field Theories in
                 Two Dimensions} & \cr
\+ &           & World Scientific, to be published & \cr
\+ &$\q{\kau}$ & H.G.\ Kausch, G.M.T.\ Watts, {\it A Study of $\w$-Algebras
                 Using Jacobi Identities}& \cr
\+ &           & Nucl.\ Phys.\ {\bf B354} (1991) p.\ 740 & \cr
\+ &$\q{\blm}$ & R.\ Blumenhagen, M.\ Flohr, A.\ Kliem, W.\ Nahm,
                 A.\ Recknagel, R.\ Varnhagen & \cr
\+ &           & {\it $\w$-Algebras with Two and Three Generators},
                 Nucl.\ Phys.\ {\bf B361} (1991) p.\ 255 & \cr
\+ &$\q{\wirrep}$
               & W.\ Eholzer, M.\ Flohr, A.\ Honecker,
                 R.\ H{\"u}bel, W.\ Nahm, R.\ Varnhagen  & \cr
\+ &           & {\it Representations of $\w$-Algebras with
                 Two Generators and New Rational Models } & \cr
\+ &           & Nucl.\ Phys.\ {\bf B383} (1992) p.\ 249 & \cr
\+ &$\q{\rva}$ & R.\ Varnhagen, {\it Characters and Representations of
                 New Fermionic $\w$-Algebras} & \cr
\+ &           & Phys.\ Lett.\ {\bf B275} (1992) p.\ 87 & \cr
\+ &$\q{\mfl}$ & M.\ Flohr, {\it $\w$-Algebras, New Rational Models and
                 Completeness of the} & \cr
\+ &           & {\it $c=1$ Classification}, preprint BONN-HE-92-08 (1992) &\cr
\+ &$\q{\twistpap}$
               & A.\ Honecker, {\it Darstellungstheorie von $\w$-Algebren und
                 Rationale Modelle in der} & \cr
\+ &           & {\it Konformen Feldtheorie},
                 Diplomarbeit BONN-IR-92-09 (1992) &\cr
\+ &           & {\it Automorphisms of $\w$-algebras and Extended Rational
                 Conformal Field Theories} & \cr
\+ &           & in preparation & \cr
\+ &$\q{\supwir}$
               & R.\ Blumenhagen, W.\ Eholzer, A.\ Honecker, R.\ H{\"u}bel &\cr
\+ &           & {\it New N=1 Extended Superconformal Algebras with Two and
                 Three Generators } & \cr
\+ &           & preprint BONN-HE-92-02 (1992),
                 to be published in Int.\ Jour.\ Mod.\ Phys.\ {\bf A} & \cr
\+ &$\q{\supwirrep}$
               & W.\ Eholzer, A.\ Honecker, R.\ H{\"u}bel & \cr
\+ &           & {\it Representations of N=1 Extended Superconformal
                 Algebras with Two Generators } & \cr
\+ &           & preprint BONN-HE-92-28 (1992) & \cr
\+ & $\q{\bouwknegt}$
               & P.\ Bouwknegt, {\it Extended Conformal Algebras from
                 Kac-Moody Algebras} & \cr
\+ &           & Proceedings of the meeting `Infinite dimensional Lie algebras
                 and Groups' & \cr
\+ &           & CIRM, Luminy, Marseille (1988) p.\ 527 & \cr
\+ &$\q{\zam}$ & A.B.\ Zamolodchikov & \cr
\+ &           & {\it Infinite Additional Symmetries in Two-Dimensional
                      Conformal Quantum Field}  & \cr
\+ &           & {\it Theory}, Theor.\ Math.\ Phys.\ 65 (1986) p.\ 1205  & \cr
\+ & $\q{\gepnera}$
               & D.\ Gepner, {\it Exactly Solvable String Compactifications
                 on Manifolds of} & \cr
\+ &           & {\it SU(N) Holonomy}, Phys.\ Lett.\ {\bf B199}
                 (1987) p.\ 380 & \cr
\+ & $\q{\gepnerb}$
               & D.\ Gepner, {\it Space-Time Supersymmetry in
                 Compactified String Theory and} & \cr
\+ &           & {\it Superconformal Models}, Nucl.\ Phys.\ {\bf B296}
                 (1988) p.\ 757 & \cr
\+ & $\q{\banks}$
               & T.\ Banks, L.J.\ Dixon, D.\ Friedan, E.\ Martinec & \cr
\+ &           & {\it Phenomenology and Conformal Field Theory Or
                 Can String Theory predict} & \cr
\+ &           & {\it the Weak Mixing Angle?},
                 Nucl.\ Phys.\ {\bf B299} (1988) p.\ 613 & \cr
\+ & $\q{\pope}$
               & H.\ Lu, C.N.\ Pope, {\it On Realisations of W Algebras} &\cr
\+ &           & preprint CTP TAMU-22/92 (1992) & \cr
\+ & $\q{\horst}$
               & H.G.\ Kausch, {\it Chiral Algebras in Conformal Field
                     Theory} & \cr
\+ &           & Ph.D.\ thesis, Cambridge University, September 1991 & \cr
\vfill
\end